\documentclass[preprint,12pt]{elsarticle}
\usepackage{lineno}
\usepackage{bm}
\usepackage{amsmath}
\usepackage{mathptmx}
\usepackage{amsfonts}
\usepackage{amssymb}
\usepackage[toc,title]{appendix}
\usepackage{multibib}
\usepackage[misc]{ifsym}
\usepackage{adjustbox}
\usepackage{multirow}
\usepackage{makecell}

\allowdisplaybreaks
\RequirePackage{mymacro}
\begin{document}
\begin{frontmatter}
\title{Exploring parametrized dark energy models in interacting scenario}

\author{Sangita Goswami}
\ead{vbsangita91@gmail.com}
\author{Sudipta Das\corref{mycorrespondingauthor}}
\cortext[mycorrespondingauthor]{Corresponding author}
\affiliation{Department of Physics, Visva-Bharati, Santiniketan -731235, India}
\ead{sudipta.das@visva-bharati.ac.in}

\begin{abstract}
In the present work, we have studied the dynamics of accelerating universe considering a simple parametrization of the equation of state parameter in an interacting scenario. In this toy model, the dark energy component is allowed to interact with the dark matter component through a source term. The expressions for various relevant cosmological parameters for the proposed parametrized model have been obtained and it has been found that the proposed model consistently drives the late time cosmic acceleration of the universe. We have also carried out the Bayesian analysis using recent observational datasets in order to obtain the best fit values of the model parameters. It has been found that the dark energy model with $Q \propto H \rho_{de}$ can provide a possible resolution of the Hubble tension in an interacting scenario where the dark energy component decays into the dark matter. 
\end{abstract}

\begin{keyword}
interaction; parametrization; equation of state parameter; Hubble parameter; cosmographic analysis
\end{keyword}
\end{frontmatter}
\section{Introduction}
The late-time cosmic speed-up is one of the most intriguing and challenging issues in cosmology today. Recent observational data like Supernova data (SNIa) \cite{riess1998observational, perlmutter1999measurements, arnaud2016planck, ahn2012ninth}, Cosmic Microwave Background  Radiation (CMBR) data \cite{jarosik2011seven}  has provided us the evidence in favour of `cosmic acceleration'. The evolution of the late universe is believed to be driven by `Dark Energy', which has a large negative pressure to counter balance the gravity and makes up about 70\% of the total energy density of the universe. Although several dark energy models have been constructed, viz, the cosmological constant model ($\Lambda$CDM), dynamical dark energy models like quintessence \cite{RatraPRD1988, CaldwellPRL1998}, phantom dark energy \cite{Caldwell_2002, CaldwellPRL2003}, Chaplygin gas model \cite{BentoPRD2002}, holographic dark energy \cite{Bousso_2002, fischler1998} etc., no single theory can be definitively considered as the best candidate for dark energy. Each of these models has its strengths and weaknesses. The cosmological constant remains the simplest and most consistent with observational data, but it also faces challenges, such as the fine-tuning problem, the coincidence problem, Hubble tension problem etc. The other theories offer alternative explanations and potential resolutions to these issues, but these models being phenomenological, they also introduce their own complexities and uncertainties. So, search is still on for a suitable dark energy candidate.\\
 Though most of the dark energy models have been framed considering a non-interacting scenario, the possibility of an interaction between dark energy and dark matter components have also been explored, which could influence the expansion rate of the universe. As nothing much is known about the nature of the dark energy, an interaction between the two components may be useful and can provide a more general scenario. 
 It has been shown by a number of authors \citep{ahn2012ninth,DiValentino:2021izs,zimdahl2004statefinder,zimdahl2012models,del2009interacting,olivares2005observational,sinha2023perturbation,das2015cosmic,nong2024} that an interacting scenario can provide solution to a number of cosmological problems, such as coincidence problem, and provides an explanation to why the dark energy domination starts during later epochs. 
In the present work we have considered an interacting scalar field dark energy model in FRW space time with a well-known parametrized form of the equation of state parameter $w_{de}$ \cite{brout2022pantheon+}. This parametrized form has been investigated and constraints on cosmological parameters of the model have been obtained from the Pantheon+ analysis of 1701 light curves of 1550 distinct Type Ia supernovae in the redshift range $0.001 < z < 2.26$. However, the proposed parametrized model could not resolve the Hubble tension between local measurements and early universe predictions. Motivated by this, we considered the parametrized form proposed by Brout et al. \cite{brout2022pantheon+} in an interacting scenario where the two dark sectors, viz., the scalar field and the matter field, are considered to interact with each other via a source term $Q$. The motivation is to check whether the proposed parametrization can resolve or at least alleviate the Hubble tension in an interacting scenario.  Efforts have been made to obtain precise solutions for various cosmological parameters as these parameters are crucial for understanding the composition, structure and evolution of the universe. The best fit values of the model parameters have been obtained by constraining them using the recent observational dataset.  \\
The paper is organized as follows: Section \ref{secequations} deals with the basic framework where we try to obtain the functional form of the Hubble parameter for the chosen parametrized form of the equation of state parameter in an interacting scenario. In section \ref{results}, we have carried out a detailed data analysis and have constrained the model parameters using Hubble, CC and Pantheon dataset. The observational results as well as the dynamical behavior of the proposed model has been presented in section \ref{results}. Finally, section \ref{conclusion} contains some concluding remarks. 

\section{Basic framework of scalar field dynamics in an interacting scenario:}\label{secequations}
 The current cosmological model is described by the spatially flat, homogeneous and isotropic Friedmann-Robertson-Walker (FRW) universe 
$$ ds^2 = dt^2 - a^2(t)\left[dr^2 + r^2 d\theta^2 + r^2 sin^2\theta d\phi^2 \right]$$
where $a(t)$ is the scale factor at time $t$ and the scale factor is normalized to $a(t_0) = 1$, the present time being denoted by $t_0$. For a universe filled  with matter and scalar field components, the Einstein's field equations are given by (in units of $8\pi G = c = 1$)
\begin{eqnarray}
  3H^2=\rho_{m} +\rho_{\phi}=\rho_{m}+\frac{1}{2}  \dot{\phi^2}+V(\phi)\label{feq1}\\
  2\dot{H} +3H^2= -p_{\phi}= -\frac{1}{2}  \dot{\phi^2} + V(\phi)\label{feq2}
\end{eqnarray} 
where $H=\frac{\dot{a}}{a}$ is the Hubble parameter. $\rho_m$, $\rho_{\phi}$ and $p_{\phi}$ are the energy density and pressure components of the different constituent sectors ($p_m=0$ for the matter sector) which are related by an equation of state (EoS) parameter $w_{\phi}=\frac{p_{\phi}}{\rho_{\phi}}$.
Along with these Einstein's equations, the matter and scalar field dark energy components will also satisfy their respective conservation equations. Usually the energy density components are considered to be non-interacting and conserved by itself such that the matter field and the dark energy components satisfy the conservation equations separately as follows :
\begin{eqnarray}
  \dot{\rho}_{m} + 3H \rho_{m}  = 0\label{fcon1}\\
  \dot{\rho}_{\phi} + 3H \rho_{\phi} (1 + w_{\phi}) = 0\label{fcon2}
  \end{eqnarray}
However, since the true nature of the mysterious `dark energy' candidate is not known, there is no a priori reason to consider these components as non-interacting. In fact, considering an interaction between the two contributing components may provide a more general scenario in this respect.  As mentioned earlier, cosmological models having an interaction between dark matter and dark energy components can also provide a solution to the coincidence problem (see \cite{Wang_2016} and the references therein). Following this line of thought, we consider that the scalar field  and the dark matter components interact with each other through a source term $Q$. The sign of $Q$ will determine the direction of energy flow between the two dark sectors. Thus for an interacting model, equations (\ref{fcon1}) and (\ref{fcon2}) gets modified as
\begin{eqnarray}
  \dot{\rho}_{m} + 3H \rho_{m}  = -Q\label{fcon3}\\
  \dot{\rho}_{\phi} + 3H \rho_{\phi} (1 + w_{\phi}) = Q\label{fcon4}
\end{eqnarray}
 The total energy density due to the dark matter ($\rho_{m}$) and dark energy ($\rho_{\phi}$) components remains conserved but one component can decay into the other.
We assume a simple functional form of Q as 
\begin{equation}
    Q=\beta H \rho_{\phi}\label{fQ}
\end{equation}
where $\beta$ is an arbitrary constant which determines the strength of the coupling. Till now there is no fundamental theory which can determine the form of the coupling term directly and usually the coupling term is chosen phenomenologically. In equation (\ref{fQ}), we have assumed this particular form of $Q$ prompted by its mathematical simplicity and also to keep the equation dimensionally correct. This is because from equations (\ref{fcon3}) or (\ref{fcon4}), one can see that left hand side terms have the dimension of the energy density multiplied by $H$ which has the dimension of $(\mathrm{time})^{-1}$ and hence equation (\ref{fQ}) becomes a natural choice. Also the sign of $\beta$ in equation (\ref{fQ}) will determine the direction of flow of energy. If $\beta < 0$, energy flows from dark energy to dark matter (DE $\rightarrow$ DM) component, while if $\beta >0$, direction of flow of energy will be from dark matter to dark energy (DM $\rightarrow$ DE) component.\\
Now  using equation (\ref{fQ}) along with the standard relations  $\frac{d}{dt} \equiv -H(1+z)\frac{d}{dz}$ and $1+z = \frac{1}{a}$, we can rewrite equations (\ref{fcon3}) and (\ref{fcon4}) in a simplified form as
\begin{eqnarray}
(1+z)\frac{d\rho_{m}}{dz}-3\rho_{m} = \beta\rho_{\phi}\label{fconr}\\
\frac{d\rho_{\phi}}{dz} = \left(\frac{3(1+w_{\phi})-\beta}{1+z}\right)\rho_{\phi}\label{fcon5}
\end{eqnarray}\\
\subsection{Cosmological solutions for a well known parametrization  of the equation of state parameter:}
Now out of the four equations (\ref{feq1}), (\ref{feq2}), (\ref{fcon3}) and (\ref{fcon5}), only three are independent and we have four unknown parameters $a$, $\rho_{m}$, $\phi$ and $V(\phi)$ to solve for. So we have the freedom to choose or assume one more constrain equation. In order to solve the system of equations, we consider a well-known parametrized form  of equation of state parameter $w_{\phi}$ as a function of red shift parameter $z$ as \cite{brout2022pantheon+}
\begin{equation}
  w_{\phi} = w_{0}+w_{1}(1+z)\label{fwde}  
\end{equation}
If the values of $w_{0}$ and $w_{1}$ are such that $w_{\phi}$ happens to be $< -\frac{1}{3}$, this parametrized model will be able drive the present observed acceleration. At $z=0$, $w_{\phi0} = w_{0}+w_{1}$ which will determine the present value of the equation of state parameter. Brout et al. \citep{brout2022pantheon+} have considered  this parametrization for non-interacting case and have obtained constraints on the cosmological parameters for various datasets. They have shown that for the combined SH0ES + Pantheon+ dataset, the present value of Hubble parameter comes out as $H_0 = 73.3 \pm 1.1 ~km/s/Mpc$ whereas for other combined datasets like Planck + Pantheon$+$ dataset or Planck + BAO + Pantheon+ dataset, the value comes out to be $H_0 = 67.4^{+1.2}_{-1.1}~km/s/Mpc$ which indicates that this particular parametrization of the EoS parameter could not explain the `Hubble tension' between local measurements and early universe predictions from the cosmological model. In the present work our objective is to check the efficacy of the parametrized form in an interacting scenario.\\
Now substituting equation (\ref{fwde}) in equation (\ref{fcon5}), one can obtain the solution of $\rho_{\phi}$ as function of $z$ as
\begin{equation}
    \rho_{\phi}(z) = \rho_{\phi0}(1+z)^{3(1+w_{0})-\beta} \exp(3w_{1}z)\label{frhophi}
\end{equation}
Now putting this expression of $\rho_{\phi}$ in equation (\ref{fconr}) one obtains 
\begin{equation}
  \frac{d\rho_m}{dz} - \frac{3 \rho_m}{(1+z)} = \frac{\beta}{(1+z)}\left[\rho_{\phi0}(1+z)^{3(1+w_{0})-\beta} \exp(3w_{1}z)\right] \label{frhodphi}
\end{equation}
The integrating factor for the differential equation comes out as $(1+z)^{-3}$, which when applied to equation (\ref{frhodphi}), one obtains the expression
\begin{equation}
\frac{d}{dz} \left(\frac{\rho_{m}}{(1+z)^3}\right) = \rho_{\phi0} \beta (1+z)^{(3(1+w_{0})-\beta-4)}exp(3w_{1}z)\label{rhomi}
\end{equation}
Here, for mathematical simplicity, we put $3(1+w_{0})-\beta-4=-1$, which immediately gives a relation between $w_{0}$ and $\beta$ as $w_{0}=\frac{\beta}{3}$. Applying this condition, the differential equation (\ref{rhomi}) reduces to a simple form given by
\begin{equation}
\frac{d}{dz} \left(\frac{\rho_{m}}{(1+z)^3}\right) = \rho_{\phi0} \beta\frac{exp(3w_{1}z)}{(1+z)}\label{rhomis}
\end{equation}
Using the standard formula for the indefinite integral $\int\frac{exp(bx)}{x} dx$ 
and neglecting higher order terms, we have obtained the final expression for $\rho_{m}$ as 
\begin{equation}
\rho_{m}(z)=\beta\rho_{\phi0}e^{-3w_{1}}\left[(1+z)^3ln(1+z)+3w_{1}(1+z)^4\right]\label{frho} 
\end{equation}
Finally putting the expression of $\rho_{m}(z)$ and $\rho_{\phi}(z)$ from equations (\ref{frho}) and (\ref{frhophi}) in the first field equation (\ref{feq1}), we have obtained the expression of Hubble parameter $H(z)$ for our model as
\begin{equation}
\begin{split}
 H^{2}(z)=H_{0}^{2}\left[(1-\Omega_{m0})\left\{\beta e^{-3w_{1}}((1+z)^{3}ln(1+z)+3w_{1}(1+z)^4)
 +(1+z)^{3}e^{3w_{1}z}\right\}\right]\label{hz}
\end{split}
\end{equation}
 The expression for $H(z)$ obtained in equation (\ref{hz}) implies a more complex expansion behavior that may provide a better fit to the observational data. In the next section we will try to constrain the parameters of the considered cosmological model, particularly $\beta$ (which will provide information regarding the strength of interaction) and $H_0$ (in order to check the viability of the model in addressing the Hubble tension), using various observational samples and consequently discuss the precise cosmographic framework for the proposed interacting model.
\section{Data analysis and results}\label{results}
Data analysis is a salient area of study in cosmology where we calculate the optimal value of the model parameters using different observational datasets. This helps us to obtain more accurate and persistent results. In this section we present a brief description of the methodology and the various observational datasets used to constrain the parameters of the model. For this analysis, we have used multiple sets of observational data, like the Hubble data \cite{solanki2021cosmic}, Cosmic Chronometer data (CC) \cite{singh2023new} and the most recent Pantheon dataset \cite{scolnic2018complete}. By fitting the parameters of the proposed model to these observational datasets, we determine their best-fit values which provides a more precise cosmological framework.\\
 We have  performed the standard Bayesian analysis \cite{padilla2021cosmological} to obtain the posterior distribution of the parameters by employing a Markov Chain Monte Carlo (MCMC) method. For this we have used the publicly available emcee library package in Python \cite{foreman2019emcee} to carry out the MCMC analysis and the GetDist package \cite{lewis2019getdist} has been used for statistical analysis.\\
 The best fit values of the parameters are obtained by minimizing the likelihood function given by
\begin{equation}
    \mathcal{L} = exp\left(\frac{-\chi^2}{2}\right)
\end{equation}
where $\chi^2$ indicates the pseudo chi-squared function \cite{padilla2021cosmological}.  The definitions of the $\chi^2$ function for Hubble/CC dataset is 
 \begin{equation}
\chi_{H/CC}^2=\sum_{i=1}^{N}\frac{[h^{obs}(z_{i})-h^{th}(z_{i})]^2}{\sigma_{H}^2(z_{i})}
\end{equation}
where $h = \frac{H(z)}{H_{0}}$ is the normalized Hubble parameter and $\sigma_{H}$ indicates the error for each data point. Here superscripts {\it obs} and {\it th} refer to the observational values and the corresponding theoretical values respectively. As mentioned earlier, in this work we have  used the 77 data points of Hubble parameter measurements in the redshift range $0.07\le z \le2.36$ \cite{singh2023new} which includes  the cosmic chronometer data as well. We have also used 57 data points of Hubble parameter measurements in the redshift range $0.07\le z \le2.36$  from \citep{solanki2021cosmic}. Among the new sets of data points, 31 points have been measured via the method of differential age (DA) and remaining 26 points through BAO and other methods. We have also used the Pantheon compilation which is basically the up-to-date collection of SNIa data in the red shift range $0.01< z <2.26$ with 1048 data points (details can be found in \citep{scolnic2018complete,peng2023pantheon}).\\ 
The definition of the $\chi^2$ function for Pantheon  dataset is given by \citep{scolnic2018complete}
\begin{equation}
 \chi^{2}_{Pantheon}=\sum_{i,j=1}^{1048}(\mu^{th}-\mu^{obs})_{i} (C_{Pantheon})^{-1}_{ij} (\mu^{th}-\mu^{obs})_{j}\\ 
\end{equation}
where  $\mu^{obs}$, $\mu^{th}$ represent the observed distance modulus and the corresponding theoretical value respectively and $(C_{Pantheon})^{-1}$ corresponds to the inverse of covariance matrix for Pantheon sample. The theoretical distance modulus can be obtained as
\begin{equation}
\mu_{th}(z)=5log_{10}\frac{d_{L}(z)}{1 Mpc}+25,
\end{equation}
 where $d_{L}(z)$ denotes the luminosity distance which can be evaluated by integrating the expression
 \begin{equation}
d_{L}(z,\theta_{s}) = (1+z)\int_{0}^{z}\frac{dz'}{H(z',\theta_{s})},
 \end{equation}
where $H(z)$ is the expression of the Hubble parameter for the cosmological model and $\theta_{s}$ is the parameter space of the cosmological model.
\subsection{Analysis of results: }
\begin{figure}[!h]
\centering
\includegraphics[width=0.85\textwidth]{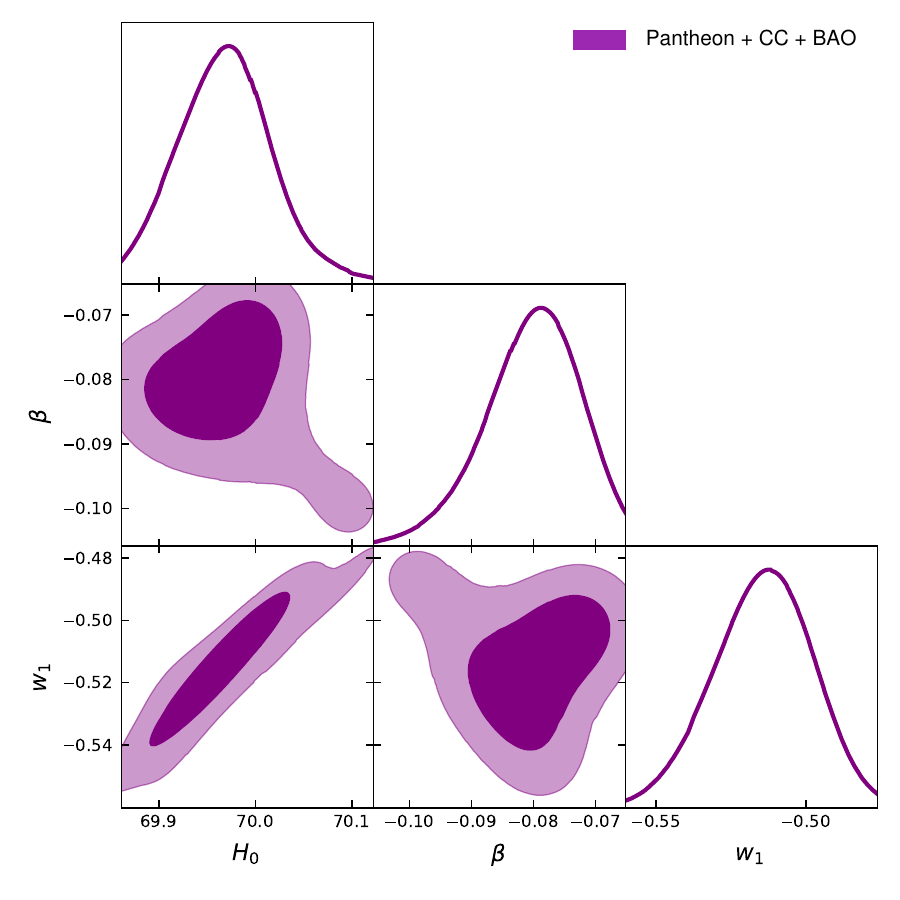}\hfill
\caption{\label{fig:tringular3} \it{1$\sigma$ - 2$\sigma$ contour plots and the marginalized constraints on the model parameters $H_{0}$, $w_{1}$, $\beta$ and for Pantheon + CC + BAO datasets. The darker and the lighter shades of the colors represent the $1\sigma$ (68.26\%) and $2\sigma$ (95.44\%) confidence levels respectively.}}
\end{figure}

In this subsection we have presented the results obtained by constraining the parameters of the proposed model using various observational datasets and their cosmological implications. For this analysis, we have set $\Omega_{m0}=0.27$ and $H_{0}$, $w_{1}$ and $\beta$ are considered as free parameters. For the MCMC analysis, we have chosen the flat priors on the base cosmological parameters as follows: the Hubble parameter $H_0 = [40, 80]~km/s/Mpc$; $w_{1} = [-1, 1]$ and $\beta =[-1, 1]$. For $H_0$, we have considered a comparatively large prior range with an aim to check whether the present parametrized interacting model can address the Hubble tension problem or not. For the parameter $\beta$, which corresponds to the strength of interaction, we have set the prior so as to accommodate both positive and negative values. This will allow us to obtain information regarding the strength of interaction as well as the direction of flow of energy. \\
\begin{table}[!h]
\centering
\begin{tabular}{|l|c|c|c|c|c|} 
\hline
Dataset & $H_{0}$ & $w_{1}$ & $\beta$  & $w_{0}=\frac{\beta}{3}$ & $w_{\phi0}$\\ 
&\small{$(km/s/Mpc)$} &  &  &  & \small{$= w_0 + w_1$} \\
\hline
$Hubble$ & $66.93^{+2.25}_{-2.11}$ & $-0.36^{+0.058}_{-0.057}$ & $-0.05^{+0.016}_{-0.026}$ & $~-0.017 $&$-0.377$\\ 
& & & & & \\
\hline
$Hubble$ & $75.6^{+1.42}_{-1.4}$ & $-0.5^{+0.032}_{-0.033}$ & $-0.018^{+0.003}_{-0.004}$ & $-0.006$&$-0.506$ \\ 
+ CC& &  &  &  &  \\
\hline
$Pantheon ~+$ & $69.97^{+0.02}_{-0.05}$ & $-0.51\pm 0.01$ & $-0.07^{+0.006}_{-0.008}$ & $-0.023$ &$-0.533$\\ 
CC + BAO& & & & & \\
\hline
\end{tabular}
\caption{\it{Best fit values of the model parameters $H_{0}$, $w_{1}$, $\beta$ and the derived parameter $w_{0}$. The last column provides the present value of the equation of state parameter}} \label{tab:parameters}
\end{table}
The best-fit values of the  model parameters $H_{0}$, $w_{1}$ and $\beta$ have been tabulated in table \ref{tab:parameters}. From these one can immediately derive the value of $w_0$ which has also been listed in table \ref{tab:parameters}. As mentioned earlier, at $z=0$, the present value of the equation of state parameter will be $w_{\phi0} = w_0 + w_1$. As evident from table \ref{tab:parameters}, for all the datasets, the present value of $w_{\phi0} < -\frac{1}{3}$  and thus the proposed model is capable of driving the current accelerated phase of expansion.  However, in an interacting scenario, the complete evolutionary picture will be portrayed by the effective EoS parameter $w_{eff}(z)$ which is defined as $w_{eff}(z)=\frac{p_{\phi}}{\rho_m + \rho_{\phi}}=\frac{p_{\phi}}{3 H^2}$. Using equation (\ref{feq2}), one can rewrite the expression for $w_{eff}(z)$ in terms of the normalized Hubble parameter $E(z)$ as 
\begin{equation}
    w_{eff}(z)=\frac{2(1+z)E\frac{dE}{dz} - 3{E}^2}{3{E}^2} = \frac{w_{\phi}\rho_{\phi}}{3 E^2}
\end{equation}
\begin{figure}[!h]
\centering
\includegraphics[width=0.75\textwidth]{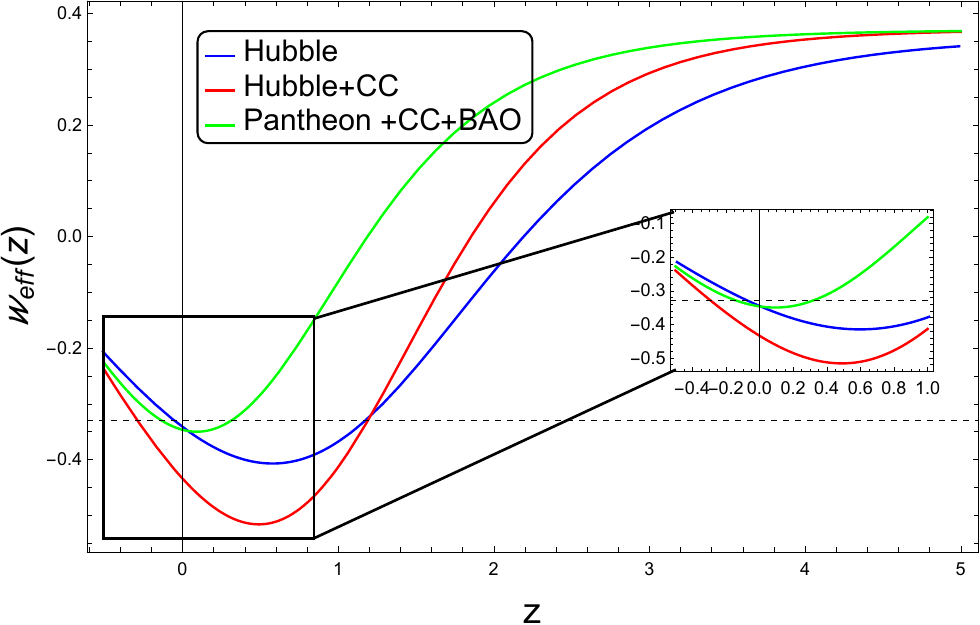}
\caption{\it{Effective equation of state (EoS) parameter $w_{eff}(z)$ for the best fit values obtained from different dataset. The inset shows the zoomed-in portion for the region $z=0$.}}
\label{fig:weff}
\end{figure}
Figure \ref{fig:weff} shows the behavior of the effective EoS parameter $w_{eff}(z)$ for the best-fit values listed in table \ref{tab:parameters}. It can be seen that for all the datasets, $w_{eff}(z)$ shows a transition from decelerating to accelerating phase and the present value is $< -\frac{1}{3}$. However, in future ($z \rightarrow -1$), $w_{eff}(z)$ again crosses the $-\frac{1}{3}$ line and this indicates that the present interacting model will have a future decelerating phase of the universe and thus can avoid the big rip problem. Thus the expansion rate of the universe will slow down at some point in the future as the influence of dark energy diminishes because of interaction. Future surveys, such as those focusing on supernovae, large-scale structure, and galaxy clustering may provide critical insights into the nature of dark energy and help determine whether future deceleration or big rip is more plausible.
\begin{figure}[!h]
\centering
\includegraphics[width= 0.63\textwidth]{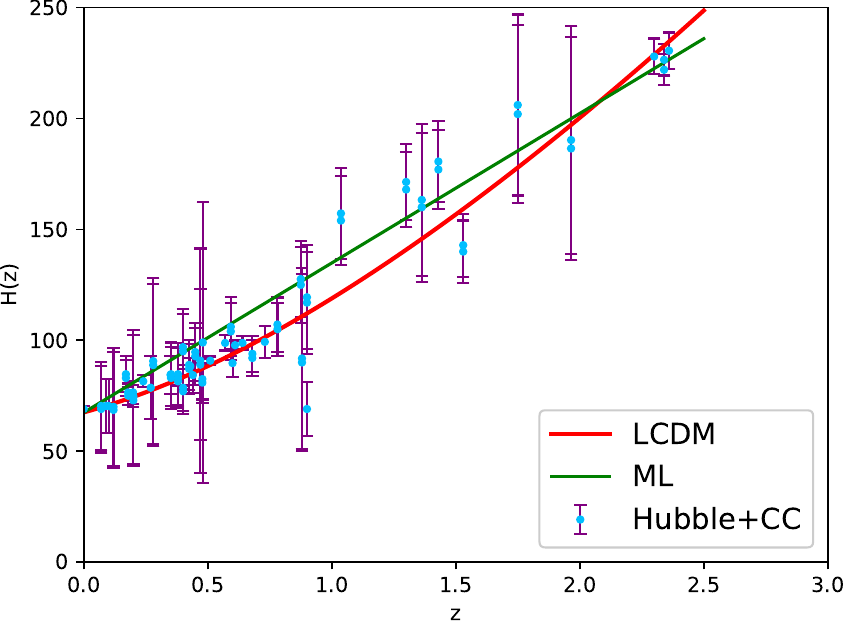}
\caption{\label{fig:Hubble2}\it{Reconstructed Hubble parameter profile for the parametrized interacting model corresponding to Hubble + CC sample. The Hubble profile for the $\Lambda$CDM model has also been plotted for comparison.}}
\end{figure}
We have also reconstructed the Hubble parameter profile for the proposed model which is shown in figure \ref{fig:Hubble2} and the profiles are found to be consistent with observational data. The blue dots with the violet bars represent the Hubble + CC sample and the green line represents the maximum likelihood plot of $H(z)$ for Hubble + CC dataset. We have also included the Hubble parameter profile for the $\Lambda$CDM model for comparison which has been shown by the red line.\\
\begin{figure}[!h]
\centering
\includegraphics[width=0.95\textwidth]{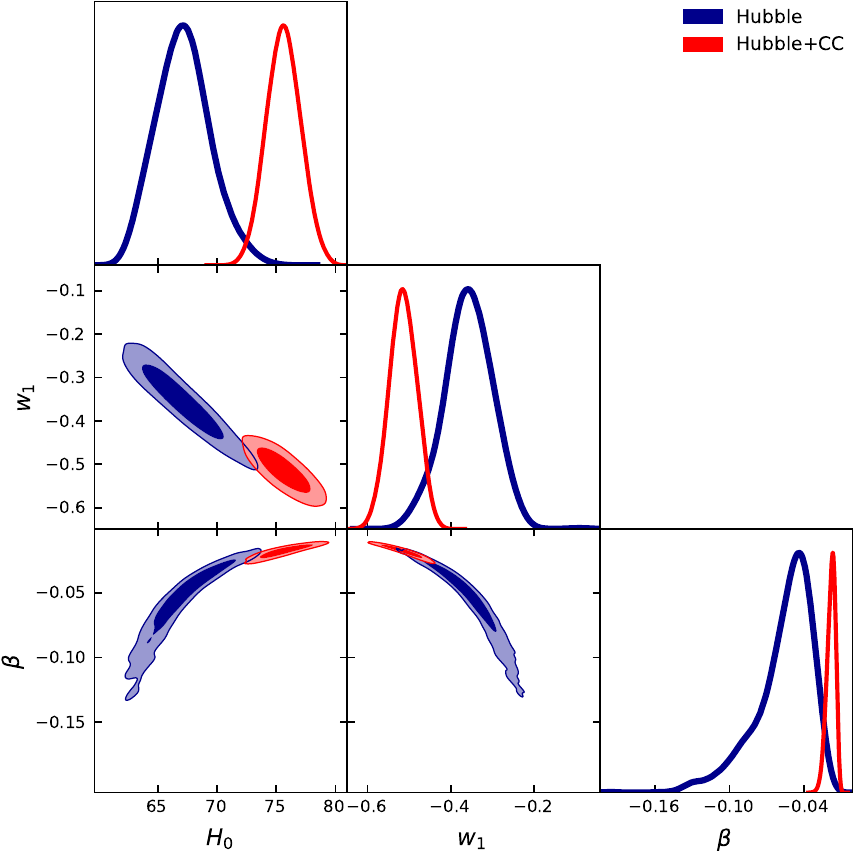}\hfill
\caption{\label{fig:tringular2} \it{1$\sigma$ - 2$\sigma$ contour plots and the marginalized constraints on the model parameters $H_{0}$, $w_{1}$ and $\beta$ for Hubble and  Hubble + CC datasets.  The darker and the lighter shades of the colors represent the $1\sigma$ (68.26\%) and $2\sigma$ (95.44\%) confidence levels respectively.}}
\end{figure}
Figure \ref{fig:tringular2} shows the $1\sigma$ and $2\sigma$ confidence level contour plot for Hubble and Hubble + CC dataset whereas figure \ref{fig:tringular3} represents the same for Pantheon + CC + BAO dataset. As evident from table\ref{tab:parameters} as well as from figure \ref{fig:tringular2}, for Hubble dataset, the value of $H_0$ comes out as $66.93~km/s/Mpc$ which is closer to the Planck measurement. But for Hubble + CC dataset, value of $H_0$ happens to be $75.6~km/s/Mpc$ which is consistent with  the SH0ES result and resolves the Hubble tension. For Pantheon + CC + BAO dataset, $H_0$ is $69.97~km/s/Mpc$ which is intermediate value and alleviates the tension partially.     

The results obtained indicate that interacting models can be effective in addressing the Hubble tension problem as these models modify the expansion rate of the universe  which can lead to a reconciliation of the $H_0$ values derived from different methods.  
\subsection{Statistical comparison with $\Lambda$CDM model:}
For a comprehensive analysis, we perform a statistical comparison of the proposed interacting dark energy model with the standard $\Lambda$CDM model. The compatibility between a given set of competing models can be determined using statistical tools like Akaike Information Criterion (AIC) and Bayesian Information Criterion (BIC) \cite{Liddle_2007}.
The AIC and BIC estimators are calculated from the maximum log-likelihood $\left({\mathcal{L}}_{max} = exp\left(\frac{-\chi^2}{2}\right)\right)$ of the model as \cite{ModAIC, Goswami:2024ymh}
\begin{eqnarray}
    AIC = -2~\mathrm{ln}~({\mathcal{L}}_{max}) + 2k  \\
BIC = -2~\mathrm{ln}~({\mathcal{L}}_{max}) + k~\mathrm{log}(N)   
\end{eqnarray}
where $N$ is the number of data points used and $k$ is the number of independent model parameters. 
For comparison with the $\Lambda$CDM model, the relative difference
$\Delta AIC = |(AIC_{model} - AIC_{\Lambda CDM})|$ and $\Delta BIC = |(BIC_{model} - BIC_{\Lambda CDM})|$ are considered. According to Jeffrey’s scale \cite{Kass1995BayesF}, if $\Delta AIC \le 2$ this will indicate that the model is strongly favoured; if it falls in the range $4 < \Delta AIC < 7$, this indicates a mild tension between the models compared; if $\Delta AIC \ge 10$, it indicates no favoured evidence and corresponds to a strong tension \cite{Mandal_2023}. For BIC analysis, $\Delta BIC < 2$ corresponds to
strong evidence in favor of the model, $2 \le \Delta BIC \le 6$ indicates moderate evidence and $\Delta BIC > 6$ shows no evidence. In this work, we have compared our proposed model with the $\Lambda$CDM model and the relative $\Delta AIC$ and $\Delta BIC$ values have been listed in table \ref{tab:AICBIC}.
\begin{table}[h!]
\centering
\begin{tabular}{|c|c|c|c|c|c|c|}
 \hline
  \hline
$Dataset$ & Model &$\chi^2_{min}$ &  AIC &$\Delta$AIC  & BIC & $\Delta$BIC\\ 
& & & & & & \\
\hline
$\multirow{2}{*}{Hubble}$ & Our model & $38.2$  &$44.2$ & $\multirow{2}{*}{1.54}$ & $43.46$ & $\multirow{2}{*}{1.29}$ \\
{\scriptsize{$(N=57)$}} & $\Lambda$CDM  & $38.66$&  $42.66$ &  & $42.17$ & \\
 \hline
     Hubble + CC & Our model &  $49.82$ &  $55.82$ &$\multirow{2}{*}{1.16}$  & $55.48$ & $\multirow{2}{*}{1.05}$ \\
 {\scriptsize{$(N=77)$}}  & $\Lambda$CDM & $50.66$  &$54.66$  
   &  & $54.43$ & \\
   \hline
  Pantheon + CC & Our model & $1025.33$ & $1031.33$  &  $\multirow{3}{*}{1.67}$ & $1034.46$ &  \\
 + BAO & $\Lambda$CDM & $1025.66$ & $1029.66$ &  & $1031.75$ & $2.71$ \\
{\scriptsize{$(N=1105)$}}& & & & & &\\
 \hline
  \hline
\end{tabular}
\caption{The $AIC$, $BIC$ values for the proposed model in comparison with $\Lambda$CDM model and the corresponding relative differences $\Delta AIC$ and $\Delta BIC$}
\label{tab:AICBIC}
\end{table}
From table \ref{tab:AICBIC}, the $\Delta AIC$ and $\Delta BIC$ values indicate that the proposed interacting model provides a very good fit to the observational datasets like Hubble, CC as compared to the $\Lambda$CDM model. For combined Pantheon data, the BIC value exhibits a very mild tension as compared to a $\Lambda$CDM model. But overall, the proposed interacting model provides a very good fit to the observational datasets and poses strong evidence compared to $\Lambda$CDM model. 

In order to quantify the tension in the determination of $H_0$, we adopt an estimator similar to \cite{Camarena:2018nbr, Roy:2023uhc}
\begin{equation}
T_{H_0} = \frac{{|H_0 - H_0^{P18}|}}{\sqrt{{\sigma^2_{H_0}} + {\sigma^2_{P18}}} }
\end{equation}
where $T_{H_0}$ represents the comparative tension in the measurement. $H_0$ and $\sigma^2_{H_0}$ represent the mean and the variance of the posterior $p(H_0)$ respectively. Similarly, $H_0^{P18}$ and $\sigma^2_{P18}$ represent the $H_0$ value corresponding to Planck 2018 measurements ($H_0^{P18}=67.4 \pm 0.5) ~km/s/Mpc$) \cite{collaboration2020planck} and $\sigma^2_{P18}$ represents the uncertainty arising from the Planck 2018 measurements \cite{collaboration2020planck}. It has been found that for the proposed interacting model, $T_{H0} \approx   5.48 \sigma$ for Hubble + CC data and $T_{H0} \approx   5.14 \sigma$ for Pantheon + CC + BAO combined dataset. So one can infer that the $\Lambda$CDM model is excluded at $5\sigma$ level.
\subsection{Cosmographic analysis:} 
Since this parametrized model described in an interacting framework could provide a resolution to the Hubble tension problem, it will be interesting and useful to study the detailed evolutionary dynamics of the model. For this we have carried out the cosmographic analysis for the proposed model. The higher order cosmographic parameters like the deceleration parameter $q(z)$, jerk parameter $j(z)$ etc. helps in checking the consistency of a cosmological model and are defined as   
\begin{eqnarray}
q(z)=-1-\frac{\dot{H}}{H^2}=\frac{(1+z)}{H}\frac{dH}{dz}-1 \label{q1}\\
  j(z) = -q(z) + 2q(z)(1 + q(z)) + (1 + z)\frac{dq}{dz}\label{eqj}
  \end{eqnarray}
\begin{figure}[!h]
\includegraphics[width= 0.49\textwidth]{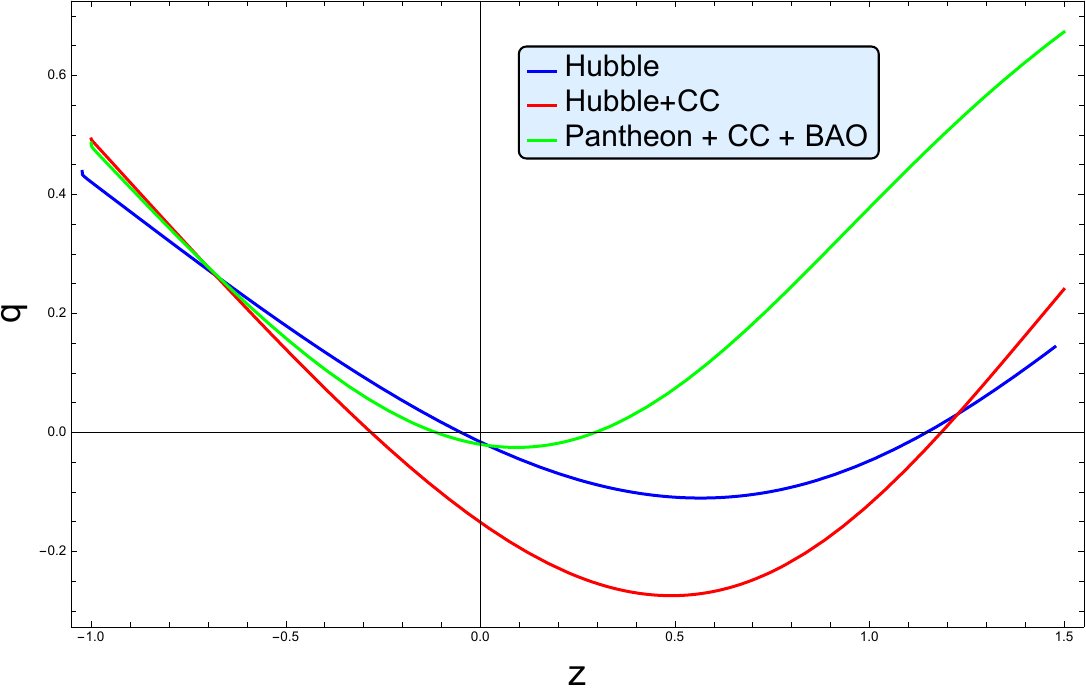}
\hfill
\includegraphics[width=0.49\textwidth]{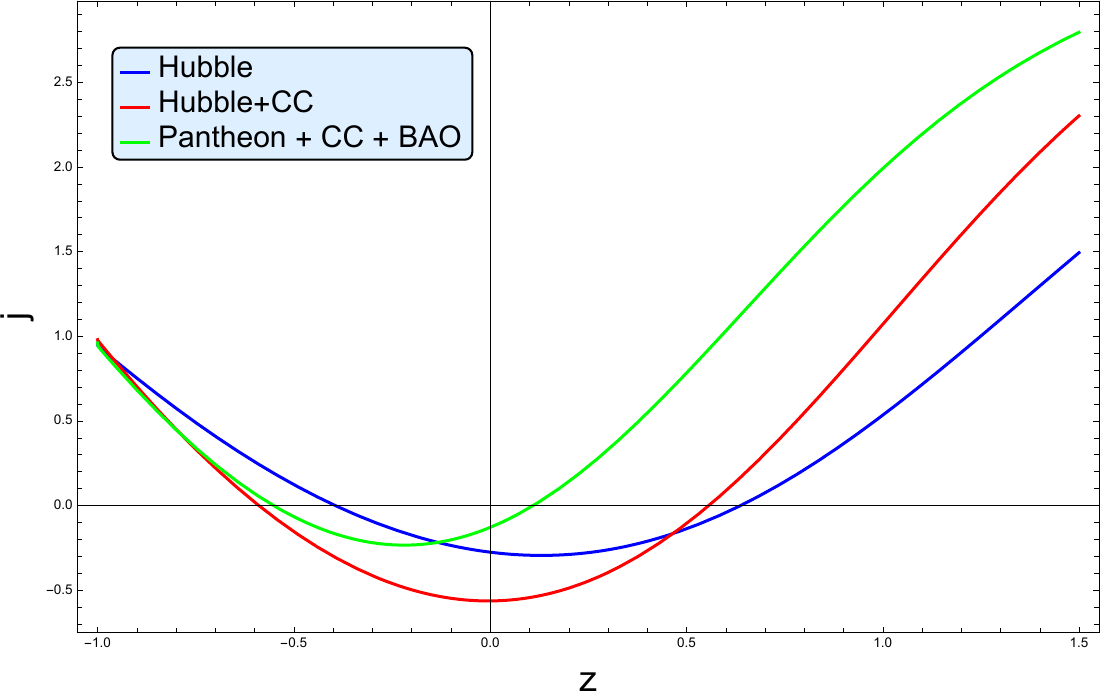}
\caption{\it{Reconstructed deceleration parameter $q(z)$ and jerk parameter $j(z)$ using the best fit values obtained from different dataset.}}
\label{fig:qj1}
\end{figure}
We have reconstructed the deceleration parameter $q(z)$ and the jerk parameter $j(z)$ using the best-fit values of the model parameters $H_{0}$, $w_{1}$ and $\beta$ listed in table \ref{tab:parameters} and the results have been shown in figure \ref{fig:qj1}. The $q(z)$ vs $z$ plot exhibits a signature flip in the deceleration parameter which indicates a late time acceleration preceded by early deceleration phase. 
The transition red shift, however, varies for different dataset; for Hubble + CC dataset, the transition occurs at $z_t > 1$ whereas for Pantheon + CC + BAO combined dataset, the transition occurs at much later time ($z_t \sim 0.3$). For all the datasets, $q(z)$ is found to re-enter a decelerating phase again which matches with the results obtained earlier. The corresponding jerk parameter for the proposed model also shows a steady decreasing nature followed by a growing trend which again indicates that the universe passes through an early deceleration phase, then transits smoothly to late time acceleration phase and approaches $\Lambda$CDM value in future.\\      
In modern cosmology $Om$ diagnostic is another useful diagnostic tool which helps in analysing the nature  of cosmological dark energy models; it is particularly helpful in testing deviations from the standard $\Lambda$CDM model. This diagnostic function was first proposed by Sahni et al. \citep{Shafieloo2008PRD}. For a spatially flat universe, $Om(z)$ is defined as
$$Om(z) = \frac{\left[\frac{H(z)}{H_0}\right]^2-1}{(1+z)^3 -1}$$ 
where $H(z)$ is the Hubble parameter of the proposed model obtained in equation (\ref{hz}). The advantage is that unlike the equation of state parameter, the $Om$ function only depends on the Hubble parameter which makes it a simpler and more direct probe of the expansion history of the universe. 
\begin{figure}[!h]
\centering
\includegraphics[width= 0.6\textwidth]{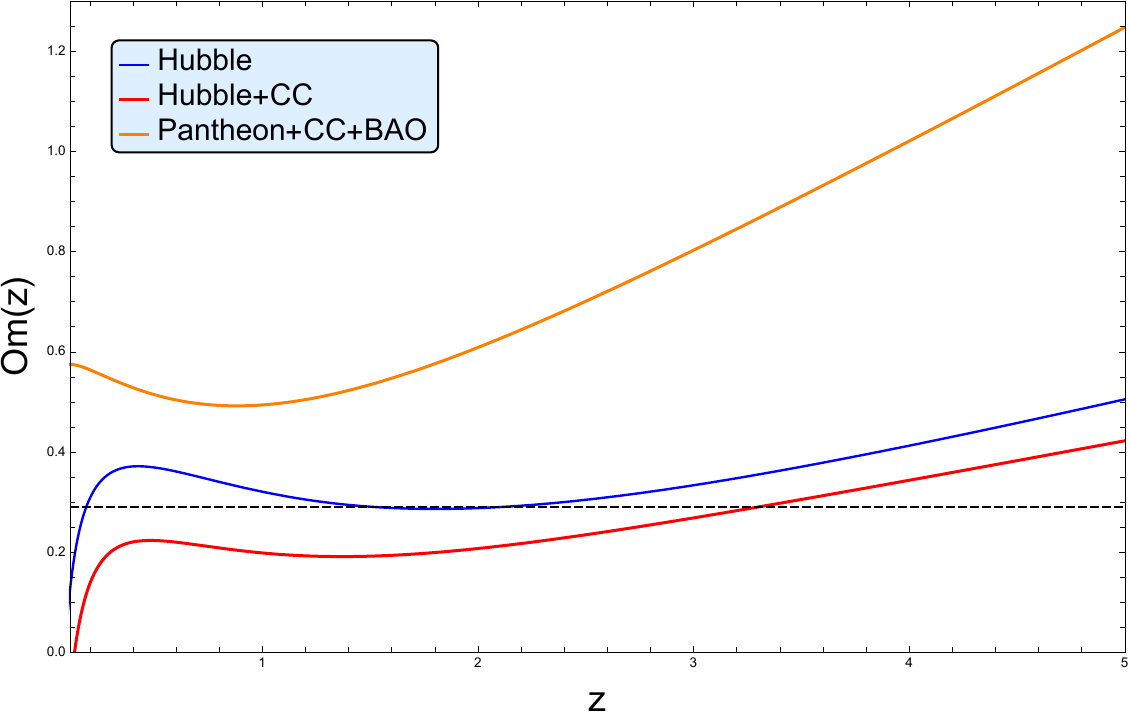}
\caption{\it{Om(z) trajectory corresponding to the best-fit values of the model parameters for the Hubble, Hubble+CC and combined Pantheon datasets.}}
\label{fig:om}
\end{figure}
The $Om(z)$ function corresponding to different observational datasets is shown in figure \ref{fig:om}. In general, a negative slope of $Om(z)$ corresponds to quintessence, a constant value corresponds to $\Lambda$CDM case (shown by dashed line in the figure) whereas a positive slope corresponds to phantom field. As evident from figure \ref{fig:om}, for Hubble and Hubble + CC dataset, $Om(z)$ trajectory shows a negative slope which indicates a quintessence behaviour of the model. Although for a small duration it shows a positive slope, which indicates a phantom behaviour, but the model mostly lies in the quintessence region. But for Pantheon + CC + BAO dataset, the $Om(z)$ characteristics shows a phantom behaviour at present.  
\subsection{Determination of sign of $Q$ :}
As observed in table \ref{tab:parameters}, the parameter $\beta$ comes out to be negative for all the datasets, which indicates that the coupling term  $Q=\beta H \rho_{\phi}$ is also negative in the present model. This in turn (from equations (\ref{fcon3}) and (\ref{fcon4})) indicates that energy is transferred from the dark energy (DE) to the dark matter (DM) sector during cosmic evolution. This is counter intuitive since one would expect the dark energy component to dominate in the later stages of cosmic evolution so as to solve for the cosmic coincidence problem. Usually interacting dark energy models are proposed with a characteristic evolution such that energy flows from the DM sector to the DE sector, ensuring a DE dominance at later times \citep{zimdahl2004statefinder,zimdahl2012models,del2009interacting, olivares2005observational}. \\ 
\begin{figure}[!h]
\centering
\includegraphics[width= 0.7\textwidth]{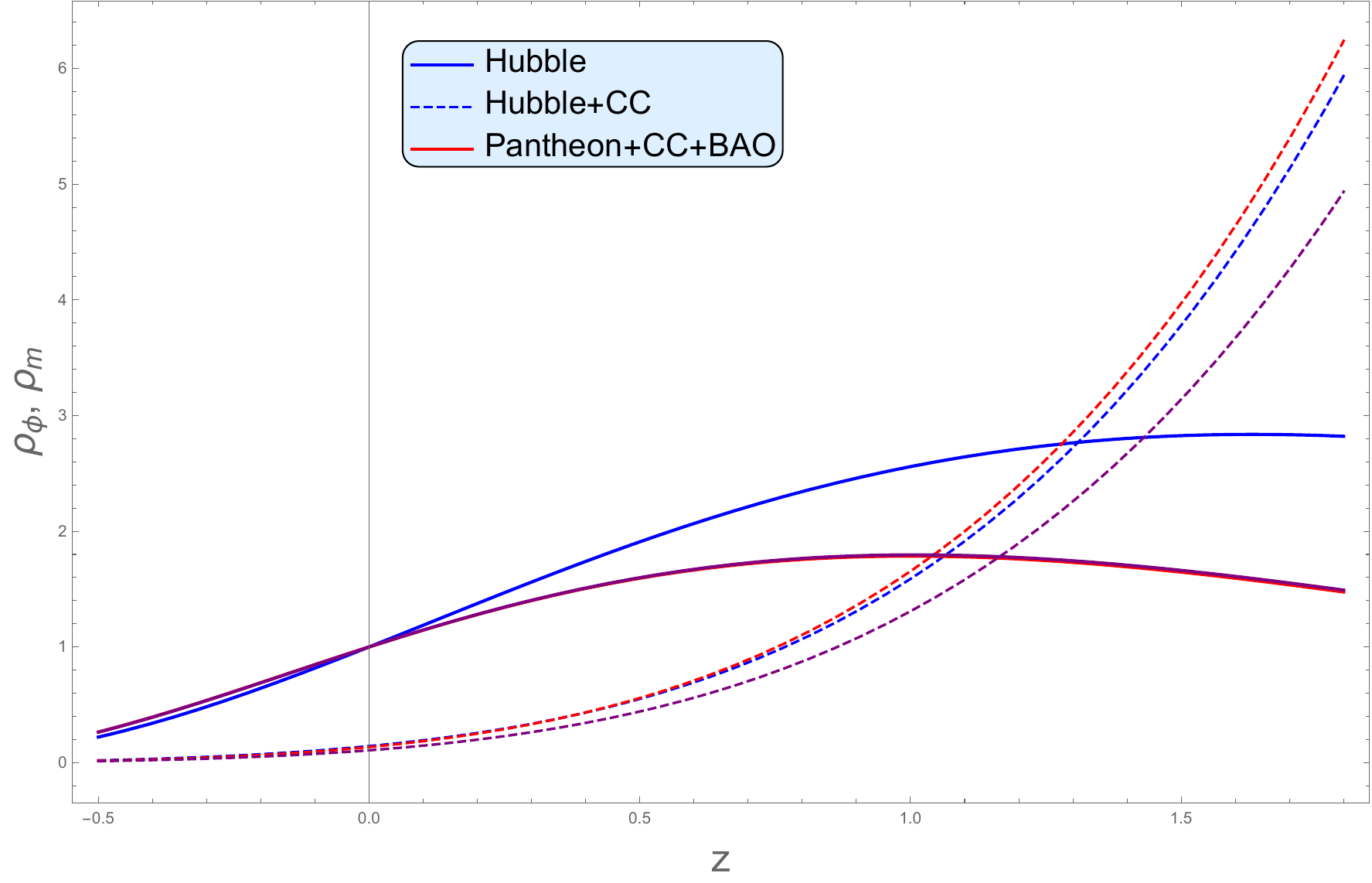}
\caption{\it{Energy densities for the dark energy component (shown by solid lines) and the matter component (shown by dotted lines) as a function of $z$ corresponding to the best fit values.}}
\label{fig:rmrphi}
\end{figure}
In the present model, however, the direction of energy flow is opposite. However, there are arguments in favour of this reverse direction of flow of energy as well. Pavon and Wang \citep{pavon2009chatelier} have shown that if dark energy is described as a fluid with a temperature close to the equilibrium, overall energy transfer should occur from the dark energy to the dark matter sector to satisfy the second law of thermodynamics and the Le Chatelier-Braun principle. Recently Li et al. \cite{li2024} have shown that in view of the latest DESI BAO data, the CMB data from Planck satellite and the SN data, interacting dark energy models with $Q \propto H\rho_{de}$ support the existence of an interaction where dark energy decays into dark matter; for other choices of the interaction term $Q$, the existence of interaction depends on the form of the proportionality coefficient. In our work also we have obtained a similar result regarding the direction of flow of energy for $Q \propto H\rho_{de}$ model, but from a different perspective. 

In order to check the viability of these reverse flow of energy, we have considered the expressions for the energy densities of the scalar field dark energy component and the matter component ($\rho_\phi$, $\rho_m$) given by equations (\ref{frhophi}) and (\ref{frho}) respectively.  Figure \ref{fig:rmrphi} represents the variation of $\rho_\phi$, $\rho_m$ as a function of $z$ for the best fit values obtained in table \ref{tab:parameters}. In the plot, the solid lines represent the energy densities for the dark energy component for different datasets whereas the dashed lines of the same colour correspond to the energy densities for the dark matter component. As seen from the figure, in spite of the fact that energy flows from the dark energy to the dark matter sector, the evolution dynamics of the universe for the present interacting model is such that the energy density of the dark energy sector is dominating over the dark matter sector at present. 

A similar evolutionary dynamics is also depicted by figure \ref{fig:weff}. The present value of $w_{\phi0}$ has been found to be $< -\frac{1}{3}$ for all the datasets and thus the scalar field dark energy component can drive the current accelerated phase of expansion. It has also been seen that $w_{eff}(z)$ re-enters a decelerating phase of expansion in future which may happen because of this reverse energy flow. This may be because of the reason that as the strength of interaction is very small and the evolution of the Hubble parameter $H(z)$ also plays an important role in the source term, the combined effect provides us with the correct cosmological scenario.
\section{Conclusion:}\label{conclusion}
In the present work we have studied the cosmological dynamics of an interacting model within the framework of a spatially flat FRW spacetime to explain the late time accelerating scenario. We have considered a well-known parametrized form of the equation of state parameter $(w_\phi)$ and have studied its dynamics by choosing a specific form for the coupling function as given in equation (\ref{fQ}). We have obtained the analytical solutions for various cosmological parameters of the model and it has been found that the cosmological parameters like the effective equation of state parameter, deceleration parameter and other cosmographic parameters show consistent evolutionary profiles. We have also performed the Bayesian analysis using recent observational datasets like Hubble data associated with Cosmic Chronometer data and Pantheon data to obtain the best fit values of the model parameters. The best fit value of $H_0$  comes out to be $75.6 ~km/s/Mpc$  for $Hubble + CC$ combined dataset which matches with the SH0ES value and thus resolves the Hubble tension. For combined Pantheon + CC + BAO dataset, value of $H_0$ comes out to be $\sim 70~ km/s/Mpc$ which is closer to the SH0ES result and resolves the tension partially. The systematic errors in distance measurements or biases in measurements of the two datasets could be the reason for this difference or inconsistency. It suggests the need for further investigation into either new physics, new systematic error sources, or more refined data that might remove the inconsistency.

It has been found that the parameter $\beta$ in the coupling term $Q$ comes out to be negative which indicates that the flow of energy due to interaction is from the dark energy sector to the dark matter sector. However despite of this reverse flow of energy, the present value of $w_{\phi0}$ has been found to be $ < -\frac{1}{3}$ for all the datasets and thus drives the current accelerated phase of expansion. So, this interacting model may provide a solution to the coincidence problem. We have also investigated the viability of these reverse flow of energy by plotting the energy densities of the scalar field dark energy
component ($\rho_{\phi}$) and the matter component ($\rho_m$) against $z$ and it has been found that despite of the reverse direction of flow of energy,  the energy density of the dark energy
sector dominates over the dark matter sector at present. Thus the evolution dynamics of the universe for the proposed interacting model provides the correct cosmological scenario. So, this interacting model has provided an insight to explain the coincidence problem as well as to resolve the Hubble tension.\\
\flushleft
{\bf{Acknowledgement}}\\
 SD acknowledges financial support from ANRF, Government of India through the core research grant CRG/2023/000185. SD would also like to acknowledge IUCAA, Pune for providing support through the associateship programme. The authors would like to acknowledge the support and facilities provided through ICARD, Pune at the Department of Physics, Visva-Bharati, Santiniketan.
 
\bibliographystyle{elsarticle-num}
\bibliography{bibliography}  
\end{document}